\documentclass{emulateapj}

\shorttitle{Spitzer Observations of LABs}
\shortauthors{Webb et al.}

\begin{document}


\title{Spitzer Observations of Extended Lyman-$\alpha$ Clouds in the SSA22 Field}


\author{T.M.A. Webb\altaffilmark{1}}
\email{webb@physics.mcgill.ca}
\author{T. Yamada\altaffilmark{2}}
\author{J.-S. Huang\altaffilmark{3}}
\author{M.L.N Ashby\altaffilmark{3}}
\author{Y. Matsuda\altaffilmark{4,5}}
\author{E. Egami\altaffilmark{6}}
\author{M. Gonzalez\altaffilmark{1}}
\author{T. Hayashimo\altaffilmark{7}}

\altaffiltext{1}{McGill Unviersity Department of Physics, 3600 rue University, Montr\'eal, QC, H2A 2T8} 
\altaffiltext{2}{Astronomical Institute, Graduate School of Science, Tohoku University, Aramaki, Aoba-ku, Sendai 980-8578, Japan}
\altaffiltext{3}{Harvard-Smithsonian Center for Astrophysics, 60 Garden Street, Cambridge, MA 02138}
\altaffiltext{4}{Department of Astronomy, Graduate School of Science, Kyoto University, Kyoto 606-8502, Japan}
\altaffiltext{5}{Optical and Infrared Astronomy Division, National Astronomical Observatory of Japan, 2-21-1 osawa, Mitaka, Tokyo 181-8588, Japan}
\altaffiltext{6}{Department of Astronomy, University of Arizona, 933 N Cherry Avenue, Rm. N204, Tucson, AZ 85721-0065}
\altaffiltext{7}{Research Center for Neutrino Science, Graduate School of Science, Tohoku Univsersity, Sendai 980-8578, Japan} 






\begin{abstract}
We present the results of a Spitzer IRAC and MIPS 24$\mu$m study of extended
Lyman-$\alpha$ clouds (or Lyman-$\alpha$ Blobs, LABs) within the SSA22 filamentary structure at $z = $ 3.09. 
 We detect 6/26 LABs in all  IRAC filters, four of which are also detected at 24$\mu$m, and find good
 correspondence with the 850$\mu$m measurements of \citet{gea05}.  An analysis of the rest-frame ultraviolet, optical, near- and mid-infrared
 colors reveals that these six systems exhibit signs of nuclear activity (AGN) and/or extreme star formation.  Notably, they have properties that bridge galaxies dominated
by star formation (Lyman-break galaxies; LBGs) and those with AGNs (LBGs classified as QSOs). The LAB systems not detected in all four IRAC bands, on the other hand, are, as a group, consistent with pure star forming systems, similar to the majority of the  LBGs within
 the filament.      These results  indicate that the galaxies within LABs do not comprise a homogeneous population, though they are also  consistent with scenarios in which
 the gas halos are ionized through a common mechanism such as galaxy-scale winds driven by the galaxies within them,   or gravitational heating of the collapsing cloud itself.

\end{abstract}





\section{Introduction}

Understanding the complex physics of galaxy formation is one of the key goals of astronomy today.  Observations at  high-redshift show
that the young universe was strikingly different than it is locally and  while many high-redshift galaxy populations have low-redshift analogues, significant
evolution is seen in their global properties such as their clustering strength \citep{coi04,lefev05}, the shape of their luminosity function \citep[e.g.][]{cir08} and their overall contribution
to the integrated background light \citep[e.g.][]{web03} and the stellar mass density of the universe \citep{rud06}.   Individually, the galaxies themselves are generally  more violent and dynamic objects than typical galaxies today.  They are more gas-rich, show higher levels of star formation and nucleic activity \citep{cha05,fin06}, experience increased rates of major mergers \citep{pat02}, and exhibit signs of energetic outflows \citep{shap03}. 
The role that these processes play in building and shaping galaxies is still poorly understood, as is the link between different evolutionary phases and the
importance of  properties such as mass and the local environment in driving evolution.   

Hierarchical structure formation models predict, and observations confirm,  that the first objects to form in the early universe will be highly clustered \citep[e.g.][]{kai84,gia98}. For this reason, many studies of galaxy evolution focus on single large structures at high-redshift which, in addition to isolating the most active  sites of structure formation at early times,   provides large samples of equidistant galaxies over a range in mass, luminosity and evolutionary stages.    Such studies of high density environments have recently detected 
a new population of extended Lyman-$\alpha$ (Ly$\alpha$) emitting nebulae at high-redshift \citep{fyn99,keel99,ste00,fran01,pal04,mat04, dey05}.  The properties of these so-called Ly$\alpha$ Blobs (LABs) indicate they are energetic phases in the formation of galaxies: they have projected physical extents of tens to several hundreds of  kpc,  Ly$\alpha$ luminosities of $\sim$ 10$^{44}$ erg s$^{-1}$,  and the Ly$\alpha$ emission is often morphologically irregular and dynamically complex.     They are prevalent in high-density regions which exhibit other signs of intense star formation and active galactic nuclei (AGN) activity, though the extent to which this is a selection effect has not been adequately explored through unbiased studies \citep[c.f.][]{pres08}.  

 The largest published sample of extended LABs to date is presented in \citet[][hereafter M04]{mat04}, within the SSA22 structure at $z\sim$ 3.09.  Originally discovered as an over-density of Lyman-break Galaxies (LBGs) \citep{ste98}, subsequent narrow-band imaging of the SSA22 field \citep{ste00} revealed two giant LABs within the concentration with sizes $>$ 100 kpc. Narrow-band imaging over a larger area by M04 assembled a sample  35 LABs  with sizes above a few tens of kpc and luminosities of 10$^{42-44}$ erg s$^{-1}$.  They are imbedded within a large ($>$ 60 Mpc) structure,  traced by the fainter and smaller Ly$\alpha$ Emitter (LAE) population \citep{haya04,mat05} consisting of at least three dynamically distinct filaments. The two brightest two LABs are located at the filamentary intersection.  

Clearly, the SSA22 structure is a concentration of evolutionary activity and the large number of LABs within it provides us with an opportunity to conduct a more systematic study these enigmatic systems than has been done before. The primary outstanding question concerns the mechanism responsible for their intense Ly$\alpha$ luminosities and super-galactic sizes. 
While an obvious explanation  is photo-ionization by an embedded source, many LABs do not contain UV-bright galaxies of sufficient luminosity (M04).  Neither are they powered by interactions between radio-jets and the ambient inter-galactic medium as they are radio-quiet,  unlike the similar Ly$\alpha$ halos surrounding high-redshift radio galaxies \citep{reu03}.    Alternative scenarios hold that these systems are powered by shock heating from galaxy-wide superwinds \citep{gea05,col06}, driven by intense starbursts or AGN \citep{keel99}, or through gravitational heating during the collapse of a large primordial cloud \citep[e.g.][]{smi07}.

Although exceptional UV and radio sources are not in general found within LABs, 
recent observations have shown that many (but not necessarily all)  LABs are associated with infrared luminous galaxies \citep{cha01,gea05,col06,bel08}.  The work of \citet{cha01} and \citet{gea05} 
detected 5 of the M04 LABs at 850$\mu$m ($>$ 3$\sigma$ significance).   It is notable that the Submillimeter Array has failed to detect the most luminous of these, LAB01 \citep[$S_{850{\mu}m}$ = 17 mJy;][]{ste00} indicating either extended submillimeter emission ($>$ 4{\arcsec}) or multiple submillimeter emitters confused within the beam \citep{mat07}, a point we will return to in \S 4.2.   Work with the MIPS camera on board Spitzer has also revealed infrared-luminous counterparts to a number of LABs including the largest currently known \citep{dey05,col06}.  These results all imply  classifications of ultra-/hyper-luminous infrared galaxies (ULIRGs/HyLIRGs), with infrared luminosities of $\sim$10$^{12-13}$ L$_\odot$. The  relationship between the infrared luminous systems and their respective LABs is unclear, but a weak correlation between the counterpart bolometric luminosity and the Ly$\alpha$ luminosity suggests that the size and luminosity of the LAB is controlled by global, galactic sized processes such as super-winds.

In this paper we present the first mid-infrared study of a large sample of LABs within the SSA22 filament  with IRAC \citep{faz04} and MIPS \citep{riek04} on board the Spitzer Space Telescope \citep{wer04}.   We use these observations, along with supplementary data from the literature, archives, and our own optical observations,  to identify the galaxies which are associated with the LABs and to investigate their nature and relation to the clouds. At $z = $ 3.09 these data sample the rest-frame near- and mid-infrared (NIR/MIR), and are sensitive to stellar emission, thermally radiating hot dust, and polycyclic aromatic hydrocarbon features (PAHs). The broad shape of the spectral energy distribution (SED) in this region provides a    useful diagnostic of the energy production mechanism within each galaxy that is complementary to the optical/UV and far-infrared observations.  

The paper is organized as follows:  \S 2 describes the observations and data analysis techniques of the mid-infrared, optical and archival X-ray data;  in Section \S 3 we present the infrared counterparts to the LABs;  and \S 4 contains a discussion of the constraints on the nature of the counterparts,  comparisons to previous studies, and implications for our understanding of the LABs and their relation to galaxy formation.

\section{Observations and Analysis}

\subsection{The IRAC and MIPS Observations}

The Spitzer IRAC (3.6, 4.5, 5.8, and 8.0 $\mu$m) and MIPS (24 $\mu$m)
data are part of two GTO programs, PID \#64 and \#30328 (PI: Giovanni
Fazio).  The former obtained single pointing data while the latter
obtained images over most of the LAE filament.  Although 70 and
160$\mu$m data were also obtained in program \#64, in this paper we
only discuss the 24 $\mu$m data.

In program \#64, a $\sim$5\arcmin$\times$5\arcmin\ area was imaged
with IRAC (AORID \#4397824) and MIPS (AORID \#4397568) with
integration times of 6400 sec/pixel and 1120 sec/pixel, respectively.
In program \#30328, an area of $\sim$375 arcmin$^{2}$ was imaged with
IRAC (AORIDs \#17599488, 17599744, 17600000, 17600256, 17600512) with
integration times ranging from 3000 sec/pixel to 7500 sec/pixel, out
of which an area of $\sim$225 arcmin$^{2}$ was coverd by all four
wavelengths to a uniform depth of 7500 sec/pixel.  This same area of
$\sim$225 arcmin$^{2}$ was also imaged with MIPS (AIRIDs \#17600768,
17601024) with an integration time of 1200 sec/pixel. 

The IRAC data were combined using MOPEX in tandem with the IRACproc
package \citep{schu06} version 4.2.1.  The pixels in the final
mosaics were chosen to be 0\farcs86 on a side so that the area
subtended by each pixel was exactly half that of the native IRAC
pixels.    The MIPS data were reduced and mosaicked with the MIPS Data Analysis
Tool (DAT) as described in Gordon et al. (2005).  The 24 $\mu$m images
were resampled and mosaicked with half of the original instrument
pixel scale (1\farcs25).

\begin{figure}
\epsscale{1.3}
\vskip -3cm
\hskip -1cm
\plotone{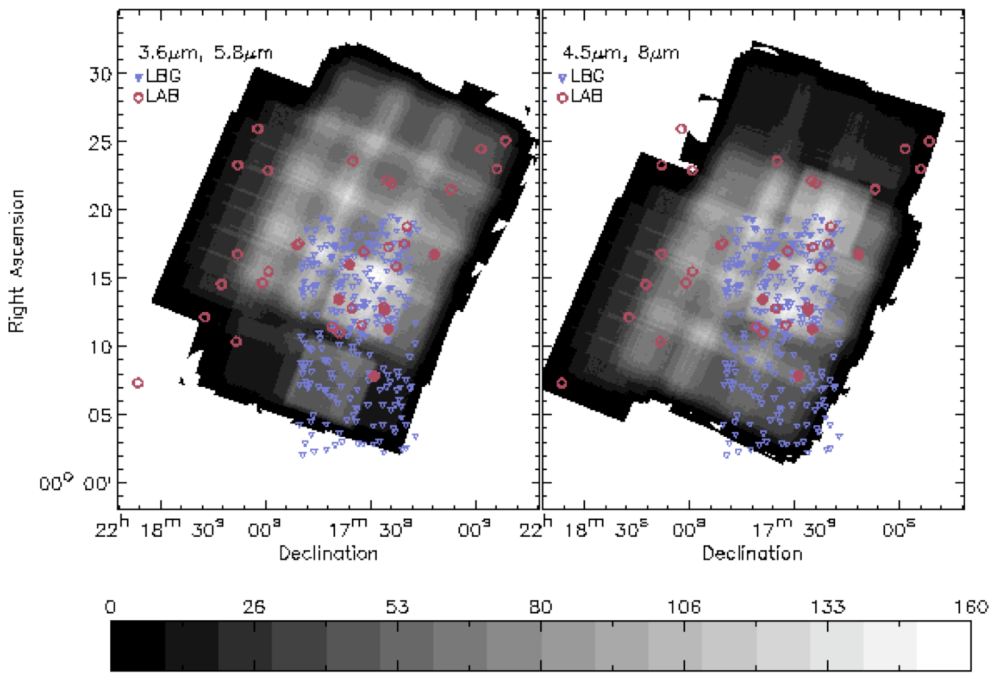}
\vskip -4.5cm
\epsscale{1.4}
\plotone{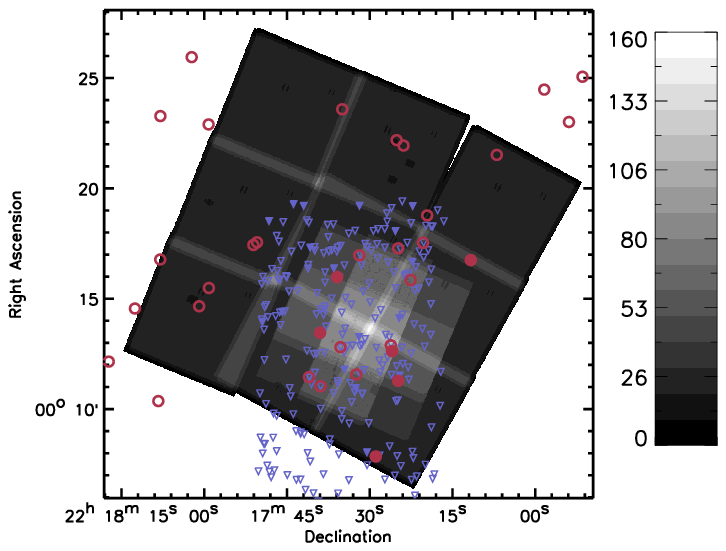}
\vskip -4cm
\caption{The Spitzer IRAC (top) and MIPS (bottom) coverage of the SSA22 field. The greyscale corresponds to the number of frames per pixel.   Also shown are the locations of the LABs in the field (red circles) and the LBGs from \citet{ste03} (blue triangles).  In both cases the filled points denote objects detected at 8.0$\mu$m. \label{field}} 
\end{figure}

Figure \ref{field} illustrates the varying IRAC and MIPS coverage of the field. 
Of the 35 LABs discovered by M04, 32 are covered by all four IRAC filters (not covered are LAB17, LAB10, LAB 28).  Of these, six are confused or contaminated by nearby bright objects (LAB03, LAB04, LAB13, LAB21, LAB30, LAB32).  This yields a working sample of 26 LABs, of which  20 lie within
the 225 arcmin$^2$ covered by MIPS and by all 4 IRAC filters to a uniform depth, except in the region of the single deep pointing.  Figure \ref{field} also shows the location of the $\sim$ 200 LBGs from \citet{ste03}, which constitute a comparison sample throughout the paper.

Photometry on the IRAC and MIPS images was performed using 3.4$''$ and 6$''$ diameter apertures respectively, and aperture corrections were applied to correct to total fluxes. Uncertainties were determined through the variance in the sky background within an aperture area, measured within an 80$\arcsec$ box surrounding each object, and the variation in depth seen in Figure \ref{field} is reflected in the uncertainty variation in Table 1. 
Table 1 lists the IRAC, MIPS  flux measurements or upper-limits for the LABs with coverage at all four IRAC filters; we also include the SCUBA measurements of \citet{cha01} and \citet{gea05} for reference.

\subsection{MMT Optical Imaging}
The SSA22 field was imaged at $r^\prime$ and $i^\prime$ (bands that approximate, but are not identical to, SDSS $r$ and $i$) during the nights of 2006 September 25, 26, and 27 with the Megacam camera McLeod et al. (2006) on the MMT atop Mt. Hopkins under conditions of stable sub-arcsecond seeing at a variety of position angles.  The 500~s 36-chip exposures were reduced using standard methods and the custom IRAF reduction package Megared. Individual exposures were first bias-subtracted and trimmed before correcting bad pixels by nearest-neighbor interpolation and flattening with evening twilight flats.  The coordinate system for each exposure was refined using the USNO B1 catalog and a priori knowledge of the relative orientations of the detector arrays; absolute accuracy in the positions of objects that are well-detected in single exposures should be better than 0.2$\arcsec$.

Using photometry of stars common to the set of exposures, illumination corrections were computed to account for the contribution of OH skyglow to the twilight flats (B. Mcleod, priv comm).  These two corrections (one at $r^\prime$, the other at $i^\prime$) were applied to all exposures in the appropriate bands.  Sky templates were then created by median-filtering the object-masked science frames; these sky templates were used to fit and subtract large-scale gradients in the background of the individual exposures on a chip-by-chip basis.
The utility SWarp was then used to project all 36 chips from each exposure to a common coordinate grid with pixels 0.32$\arcsec$ on a side (four times the Megacam pixel size) and simultaneously to fit and remove large-scale ($>$82$\arcsec$) structure from the backgrounds. Using coextensive SDSS photometry, the individual exposures were individually flux-calibrated and then correspondingly scaled/weighted when subsequently coadded to construct
the final mosaics.  The approximate 5-$\sigma$ depths of the final mosaics, after removal of artifacts, are roughly 27.1 and 25.9 AB mag at $r^\prime$ and $i^\prime$ respectively.

\subsection{Archival X-ray Data}
\subsection{Chandra}
The SSA22 field was observed with Chandra on 2001 July 10. After downloading from the archive, the data were reduced and analyzed
using standard recipes with the CIAO package v3.3.0. The observation was 
performed continuously with no significant increase in background and a 
total livetime of 77.8~ks.
The sources associated with the SSA22 LABs fall throughout the ACIS-S3 and ACIS-S2
chips. Exposure maps were computed for both chips and the resulting
fluxes corrected accordingly. 

\subsection{XMM$-$Newton}
 XMM observations were taken on 2002 November 10 and 2002 December
15. After downloading from the archive, the data were reduced and analyzed using standard techniques and SAS
v.7.0.0. Periods of high background were excluded from the data. During both 
observations, the EPIC MOS1/MOS2 and PN detectors were operated in full-window 
mode using the thin filter. In order to increase the signal to noise, we combined the 
individual MOS1+2 (i.e., one final MOS) and PN dataset for each observation. The 
total observing times were 256~ks and 110~ks for the MOS and PN detectors, 
respectively. Exposure maps were averaged and the resulting fluxes corrected 
accordingly.

\subsection{X-ray Analysis}

The Chandra and XMM datasets were searched for LAB counterparts using source
detection algorithms such as {\tt celldetect}, {\tt wavedetect} and {\tt ewavelet}.  We detect two systems:  LAB02
which has a previously published X-ray detection \citep{basu04},  is marginally detected by our 
analysis; and we detect, for the first time, LAB14. For the undetected LABs we calculated 3$\sigma$ upper limits on their fluxes derived using background
levels close to the source positions.  We have also clearly detected the two previously known QSO LBGs in the field from \citet{ste03}. 

When available, Chandra and XMM provided similar flux 
limits; this is consistent with the different effective areas, exposure times, 
background and point-spread-function of each telescope and instrument. To compute the fluxes we have 
assumed a galactic absorption of $N_H=4.8\times$10$^{20}$ cm$^{-2}$ and an intrinsic photon energy power-law slope of $\Gamma=1$.

\section{Results}
\subsection{Infrared Counterparts to LABs} 

\begin{deluxetable*}{lrrrrrrrr}
\tabletypesize{\scriptsize}
\tablecaption{Infrared Measurements or Limits for the Counterparts to LABs}
\tablewidth{0pt}
\tablehead{
\colhead{Name}  & \colhead{RA} & \colhead{Dec} & \colhead{3.6$\mu$m ($\mu$Jy)}  & \colhead{4.5$\mu$m ($\mu$Jy)} & \colhead{5.8${\mu}$m ($\mu$Jy)} &\colhead{8.0$\mu$m ($\mu$Jy)}  & \colhead{24$\mu$m ($\mu$Jy)} & \colhead{850$\mu$m (mJy)\tablenotemark{a}} }
\startdata
LAB01-a & 22:17:26.0 & 00:12:36.6  & 7.3 $\pm$ 0.1 & 9.2 $\pm$ 0.2 & 11.5 $\pm$ 0.8 & 14.2 $\pm$ 1.0  & 51.3 $\pm$ 7.5 &  16.8\tablenotemark{b} $\pm$ 2.9\\
LAB01-b & 22:17:26.1  & 00:12:32.5  & 8.4 $\pm$ 0.1 & 11.1 $\pm$ 0.2 & 14.7 $\pm$ 0.8 & 15.9 $\pm$ 1.0 & 57.5 $\pm$ 7.6 &  ... \\
LAB02-a & 22:17:39.3 & 00:13:22.0 & 5.6 $\pm$ 0.2 & 6.7 $\pm$ 0.2 & 8.3 $\pm$ 0.9 & 6.5 $\pm$ 1.4 &  $<$ 45  & $<$ 3.6  \\ 
LAB02-b & 22:17:39.1 & 00:13:30.7 & 5.8 $\pm$ 0.2 & 7.8 $\pm$ 0.2 & 10.5 $\pm$ 0.9 & 15.4 $\pm$ 1.4 & $<$ 45 & $<$ 3.6 \\ 
LAB05 & 22:17:11.7 & 00:16:44.3 & 7.7 $\pm$ 0.1 & 9.4 $\pm$ 0.2 & 10.1 $\pm$ 1.0 & 12.4 $\pm$ 1.3 &  $<$ 40 &5.2 $\pm$ 1.4 \\
LAB14 & 22:17:35.9 & 00:14:58.8 & 7.4 $\pm$ 0.1 & 10.6 $\pm$ 0.2 & 15.0 $\pm$ 0.8 & 19.6 $\pm$ 1.1 & 71.3 $\pm$ 9.8 &   4.9 $\pm$ 1.3  \\
LAB16 & 22:17:25.9 & 00:11:17.3 & 5.8 $\pm$ 0.1 & 7.3 $\pm$ 0.2 & 9.6 $\pm$ 1.0 & 15.2 $\pm$ 1.5 & 84.3 $\pm$ 9.6 &  $<$ 16.0 \\
LAB18-a & 22:17:29.0 & 00:07:50.2 & 7.3 $\pm$ 0.2 & 8.7 $\pm$ 0.3 & 15.7 $\pm$ 1.5 & 19.2  $\pm$ 1.6 & 123.8 $\pm$ 10.6 &  11.0 $\pm$ 1.5 \\
LAB18-b & 22:17:29.0 & 00:07:43.2 & 4.9 $\pm$ 0.2 & 7.8 $\pm$ 0.3 & 8.6 $\pm$ 1.5 & 17.6 $\pm$ 1.6 & $<$ 55 &  ...  \\ \\ \hline \\
LAB06 & 22:16:51.4 & 00:25:03.6& 5.2 $\pm$ 0.3 & 5.4 $\pm$ 0.8  & $<$ 9.0  & $<$ 25.0 & ... & $<$ 9.0  \\
LAB07 & 22:17:41.0& 00:11:26.1 & 2.8 $\pm$ 0.4 & 2.8 $\pm$ 0.3 & $< $ 7.0  & $<$ 10.0 & $< $ 53 &  $<$ 8.0 \\
LAB08 & 22:17:26.2 & 00:12:55.2 & 2.6 $\pm$ 0.1 & 1.0 $\pm$ 0.2  & $<$ 4.0  & $<$ 5.0  & $<$ 38 &  $<$ 27 \\
LAB09 & 22:17:51.2 & 00:17:28.0  & 2.3 $\pm$ 0.2 & 2.2 $\pm$ 0.2 & $<$ 5.0 & $<$ 10.5  & $<$ 56 & $<$ 27 \\
LAB11 & 22:17:20.3 & 00:17:32.2 & 3.0 $\pm$ 0.2 & 2.6 $\pm$ 0.2   & $<$ 5.0  & $<$ 6.0  &  $<$ 42 & $<$ 27  \\
LAB12 & 22:17:31.7 & 00:16:57.5 & 2.6 $\pm$ 0.2 & 3.0 $\pm$ 0.2 & 4.8 $\pm$ 0.9 &   $<$  6.0 & $<$ 52  & $<$ 8.0 \\
LAB15 & 22:18:08.4 & 00:10:22.5 & 1.7 $\pm$ 0.3 & 1.8 $\pm$ 0.3  & $<$ 12.0  & $<$ 8.0   & ... &  ... \\
LAB19 & 22:17:19.6 & 00:18:44.7 & 2.8 $\pm$ 0.1 & 2.4 $\pm$ 0.2 & $<$ 4.0  & $ < $ 5.5 & $<$ 49  & $<$ 27  \\
LAB20 & 22:17:35.4 & 00:12:47.6 & 1.3 $\pm$ 0.2 & $<$ 1.0  & $<$ 4.0  & $<$ 7.5 & $<$ 44 &  $<$ 8.0 \\
LAB22 & 22:17:35.0 & 00:23:34.4 & $<$ 3.0 & $< $ 2.0 & $<$ 6.0  & $<$  8.0 & $<$ 52 & ... \\
LAB23 & 22:18:08.0 & 00:23:15.5 & $ < $ 1.5 & $<$ 4.0  & $<$ 13.0  & $<$ 27 & ... & ... \\
LAB24 & 22:18:00.9 & 00:14:40.0  & $ < $ 0.8  & $< $ 1.0  & $<$ 5.0  &  $<$ 7.0 & $<$ 62 & ... \\
LAB25 & 22:17:22.6 & 00:15:51.6 & 1.2 $\pm$ 0.1 & 2.4 & $<$ 4.0  & $<$ 8.0  & $<$ 46 &  $<$ 27 \\
LAB26 & 22:17:50.4 & 00:17:33.3 & $<$ 1.0 & $<$ 1.0 & $<$ 5.0 & $< $ 10.0 & $<$ 57 & $<$ 27 \\
LAB29 & 22:16:53.9 & 00:23:00.9 & 3.6 $\pm$ 0.3  & $< $ 4.0 & $<$ 8.0   & $<$ 24.0   & ... &  ... \\
LAB31  & 22:17:39.0 & 00:11:02.4  & 1.1 $\pm$ 0.3  & $<$ 1.5 &  $<$ 6.0 & $<$ 8.0 & $<$ 50 & $<$ 27 \\
LAB33  & 22:18:12.0 & 00:14:32.6  & 1.8 $\pm$ 0.2  & 2.1 $\pm$ 0.3 & $<$ 10.0 & $<$ 10.0 & ...  & $<$ 8.0 \\
LAB34  & 22:16:58.4 & 00:24:29.2  & 1.8 $\pm$ 0.2  & $< $ 2.0 & $<$ 6.5 & $<$ 13.5 & ... & ... \\
LAB35 & 22:17:24.8 & 00:17:17.0 & 1.4 $\pm$ 0.2& 1.4 $\pm$ 0.2 & $<$ 4.5  & $<$ 6.0 & $<$ 45 &  $<$ 27 \\
\enddata
\tablenotetext{a}{From \citet{gea05} and \citet{cha01}}
\tablenotetext{b}{Because of the large uncertainty in the submm position the submm counterpart may be physically associated with either or both of the 8$\mu$m sources, though here we list it only once.} 
\end{deluxetable*}

We detect counterparts at  $>$ 5$\sigma$ in all four IRAC filters for 6/26 LABs, where the term counterpart refers to objects found within the extent of the Ly$\alpha$ emission. The 4-band detected LABs are: LAB01, LAB02, LAB05, LAB14, LAB16, and LAB18 (maintaining the M04 nomenclature).  All six lie within the MIPS image and and all but two (LAB02; LAB05) also show 24$\mu$m emission; there is a tentative 24$\mu$m detection of LAB02  but it is at $< $ 5$\sigma$  and appears spatially offset from  the 8$\mu$m detection.

\begin{figure}
\epsscale{1.75}
\plotone{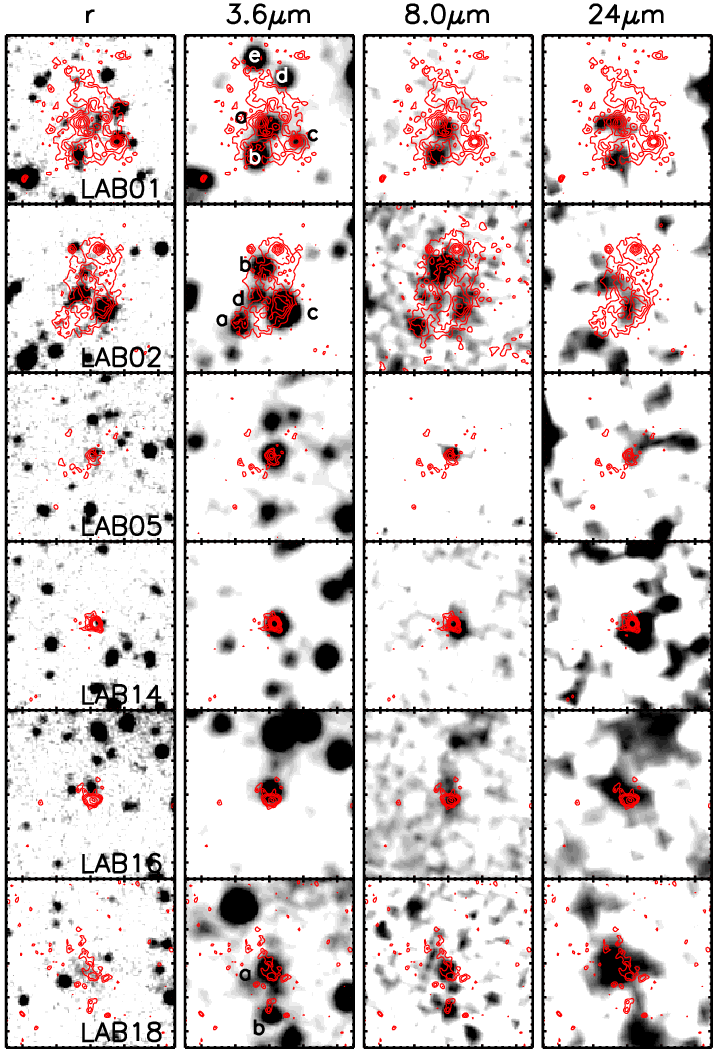}
\caption{Postage stamps of the 8$\mu$m detected LABs at $r^\prime$, 3.6$\mu$m, 8.0$\mu$m, and 24$\mu$m with the Lyman-$\alpha$ contours overlaid in red \citep{mat04}. North is up and east to the left.  The distance between large tick marks corresponds to 5$\arcsec$, and each image is 25$\arcsec \times$25$\arcsec$ in size.\label{overlay}}
\end{figure}

  Figure \ref{overlay} shows the $r^\prime$, 3.6$\mu$m, 8.0$\mu$m, and 24$\mu$m images of the  6 LABs, overlaid with Ly$\alpha$ contours from the M04 survey.  The LABs can be divided into two distinct groups:  LABs with sizes $>$ 100 arcsec$^2$ which have more than one galaxy within the extent of the Ly$\alpha$ emission (LAB01, LAB02, LAB18); and the smaller, $<$ 100 arcsec$^2$ LABs which are coincident with only one possible counterpart (LAB05, LAB14, LAB16). Below we discuss the specifics of each source.

\noindent {\bf LAB01:}  With a projected physical size of ~150 kpc, this is the largest LAB within the SSA22 filament and indeed is one of the largest  discovered to date (see also \citet{dey05}). It has been the focus of a number of studies \citep{ste00,cha01,bow04,gea07}.  
Here we briefly summarize the relevant points and refer the reader to the earlier work for details.

Multiple systems have been detected within the large extent of the Ly$\alpha$ emission including a strong submillimeter emitter ($S_{850{\mu}m}\sim$ 17 mJy) and a radio source at 1.4 GHz, \citep{cha01}); these may be the same object, though the uncertainty on the submillimeter position is substantial and is spatially consistent with other counterparts.   Deep HST STIS imaging by \citet{cha01} reveals four faint optical components in the central region of LAB01 and we recover these structures, with much lower spatial resolution, in the $r^\prime$-band MMT image (Figure \ref{overlay}).  One of these (J2 in Chapman et al.) is an Extremely Red Object and is a second possible counterpart to the submillimeter source, but is offset from the radio position.

\citet{gea07} investigated the mid-infrared properties (rest-frame NIR) of LAB01 (and LAB02)  using a subset of the data presented here.  For clarity we have adopted the counterpart labels of that paper  in Figure \ref{overlay}, for LAB01 and LAB02. There are five objects detected at 3.6$\mu$m within the extent of the Ly$\alpha$ emission which Geach et al.~have denoted {\bf a},{\bf b},{\bf c},{\bf d}, {\bf e}.  Sources {\bf e} and {\bf d} are only detected at 3.6$\mu$m  and 4.5$\mu$m and have  blue colors ($S_{3.6{\mu}m} > S_{4.5{\mu}m}$) consistent with $z<$ 1 galaxies.  Source {\bf c} is the LBG C11 \citep{ste00} which is a member of the filament and therefore possibly physically associated with the LAB , though as noted by Geach et al.~and \citet{bow04} it appears kinematically distinct from the Ly$\alpha$ emission.

Sources {\bf a} and {\bf b} are detected in all four IRAC bands as well as MIPS 24$\mu$m.  LAB01-a is located in the central region of the cloud, close to a peak in the Ly$\alpha$ emission, but its counterpart at shorter wavelengths remains unclear. For the remainder of the paper we have adopted the ERO mentioned above  as it is the closest object to the 3.6$\mu$m and 24$\mu$m centroids and is optically red, though a second possibility  is the radio source. We note however that the similar $r^\prime$-band flux of the ERO and radio source means none of our key results are strongly dependent on this assumption.

LAB01-b has not received much attention thus far, perhaps because it lies in the outer region of the Ly$\alpha$ cloud and $\sim$7{\arcsec} from the submillimeter centroid. The optical and near-infrared imaging reveals two objects separated by 2$''$, one of which is substantially brighter than the other and coincident with the center of the 3.6$\mu$m emission; hereafter we take this brighter system to be the IRAC source.

The IRAC-MIPS photometry of LAB01-a and -b are very similar: both are detected in all 4 IRAC filters and MIPS-24$\mu$m at roughly the same flux levels.     As discussed in \S 4, their infrared colors are indicative of $z\sim$ 3 galaxies and we here conclude that both systems are associated with the SSA22 filament and LAB01. As already noted by Geach et al., the projected separation is $\sim$30 kpc and hence this may be an interacting system within the LAB, though no clear evidence  of this is seen in the deep optical imaging.  

 LAB01 has now been observed with a number of different submillimeter facilities. The original SCUBA detection  \citep{cha01} is the strongest and most significant at S$_{850{\mu}m}$ = 16.8 $\pm$ 2.9 mJy,  but  the large beam (FWHM $\sim$ 15{\arcsec}) of the JCMT at 850$\mu$m  makes the positional uncertainty significant, and it is  possible that more than one source within the LAB is contributing to the submillimeter flux.  Such a scenario would explain both the very high submillimeter emission measured by the JCMT and the recent non-detection of this LAB with the Submillimeter Array (SMA) \citep{mat07}.  The SMA observations should have  detected the source, unless it was extended beyond 4{\arcsec} or composed of multiple fainter systems, confused within the JCMT beam, in which case the emission would be resolved by the SMA below the detection threshold. The 24$\mu$m detections of both LAB01-a and -b mark them both as strong infrared emitters and thus possible submillimeter sources, and their separation of $\sim$5 arcsec is certainly well below the resolution limit of SCUBA at 850$\mu$m.  
 
 This picture is confused however by two other data sets, which call into question the submillimeter measurements. The OVRO CO(4-3) observations \citep{cha04}  detected a single object 2{\arcsec} N   of the ERO, which implies only a single submillimeter emitter, however  the detection is only at a 3.2$\sigma$ significance and thus a second source of similar flux could lie just below the detection limit of these data.  Recent observations taken with the AzTec submillimeter array on the Atacama Submillimeter Telescope Experiment  at 1mm \citep{koh08} however failed to detect any submillimeter emission, to a 5$\sigma$ limit of $\sim$10 mJy \citep{koh08}.  Given the large beam-size  of $\sim$38{\arcsec} at 1mm, the existence of multiple submillimeter emitters cannot explain this non-detection.
   
\noindent{\bf LAB02:}   LAB02 is the second of three LABs in this sample with multiple objects within its large extent.  Although  this object has been identified as a possible submillimeter emitter,  the 850$\mu$m detection \citep{cha01} is of significance (2.75$\sigma$)  and we consider this an unreliable detection. As first reported by \citet{gea07}, and seen in Fig. \ref{overlay}, there are a number of systems detected in the IRAC imaging with very different observed optical and infrared colors.  The brightest object in the group, object {\bf c} in Geach et al., is a bright radio source and has  $S_{5.8{\mu}m}/S_{3.6{\mu}m} <$ 1 and $S_{8.0{\mu}m}/S_{4.6{\mu}m} <$ 1, consistent with a foreground galaxy at $z<$ 1.  Source {\bf d} (not discussed in Geach et al.) is also consistent with a low redshift galaxy: it is detected in the optical through to 4.5$\mu$m, with $S_{4.5{\mu}m}/S_{3.6{\mu}m} <$ 1, but is not visible at 5.8$\mu$m and 8.0$\mu$m.   The most likely physical counterparts to the LAB are therefore sources {\bf a} and {\bf b}. LAB02-a however is detected with a significance of only 4$\sigma$ at 8.0$\mu$m and thus its color, while consistent with $z\sim$ 3,  is poorly constrained; for that reason although its photometry is listed in Table 1, it is considered an 8$\mu$m non-detection for the remainder of the analysis. 
 
Source LAB02-b lies in the outer region of the blob. It is not detected at 24$\mu$m but has been associated with a  Chandra X-ray source \citep{basu04}.
  Using archival data, we find a 0.3--10 keV flux level at its coordinates of 
(2.5$\pm$ 1.3)$\times$10$^{-15}$ erg s$^{-1}$ cm$^{-2}$ using the Chandra data and a 
power-law spectrum, consistent with the reported values of \cite{basu04}. However, we 
find that it has a significance of $\sim$2$\sigma$ in these Chandra data (the source
is not picked up as significant in our detection algorithms) and it is {\it not} 
detected in the XMM datasets. Thus, we regard the X-ray detection as tentative:  it is likely very faint and/or slightly extended so that the larger XMM PSF
precludes a detection there. In the latter case, a deeper Chandra observations should
clarify this issue.

\noindent{\bf LAB05:} A  single counterpart exists for this LAB. It is detected at all IRAC wavelengths, at $r^\prime$ and $i^\prime$, at 850$\mu$m,  but not at 24$\mu$m.  

\noindent {\bf LAB14:} LAB14 has only one possible counterpart and this is offset from the center of the LAB by $\sim$1$''$ in all bands.  It is strongly detected at all wavelengths, from $r^\prime$ to 850$\mu$m.  The archival X-ray data have also revealed an X-ray source coincident with the IRAC position, heretofore unpublished. 
 The object was  detected with significances of $\sim$2.5$\sigma$ in {\it both} ACIS and
MOS instruments (unfortunately it fell in between chip gaps in the PN instrument)
and we thus consider it a very solid detection with a 0.3-10 keV flux level  of 2.1$\pm$0.8 $\times$ 10$^{-15}$ erg/s/cm$^2$ 

\noindent{\bf LAB16:} A counterpart is detected in all four IRAC filters, as well as MIPS 24$\mu$m.  
The peak Ly$\alpha$ flux is offset from the $r^\prime$ and 3.6$\mu$m/4.5$\mu$m counterpart by $\sim$2$\arcsec$; this may also be the case at the longer wavelengths, $>$8.0$\mu$m, but the centroid poorly determined because of the lower S/N and the larger PSF. 

\noindent{\bf LAB18:} This LAB is detected at all infrared wavelengths, though is slightly confused by a foreground neighbor at 24$\mu$m. 
 LAB18 has no obvious single $r^\prime$-band counterpart, but diffuse emission along the extent of the LAB is clearly visible.  In Figure \ref{overlay} we see
 two additional objects south of the main LAB emission which are detected at all four IRAC wavelengths, though not in the $r^\prime$ image, nor at 24$\mu$m.  
The southern-most object is coincident with another Ly$\alpha$ peak and has IRAC colors similar to the main LAB counterpart and consistent with a $z\sim$ 3 galaxy (see \S 4.1).  The fainter, middle source however, shows no Ly$\alpha$ flux and its color is contaminated by the main LAB counterpart, and we therefore have not included it as an LAB counterpart, although these three object may form a single interacting system.   This LAB has a submillimeter detection of $S_{850{\mu}m}$ = 11 mJy, and with a separation of 7{\arcsec}, both IRAC sources could contribute to the measured submillimeter flux, as may be the case with LAB01. 

\section{Discussion}

\subsection{Observed Mid-Infrared Colors: Constraints on the Nature of the   LAB Counterparts}

At $z\sim$ 3 the IRAC filters sample the rest-frame NIR. In this region pure stellar populations 
are characterized by the stellar bump at  1.6$\mu$m, with decreasing emission toward longer wavelengths.  If cold dust is present within a galaxy, band emission from PAHs becomes important over the range of $\sim$ 6 - 11 $\mu$m while  warm dust  produces emission that continues to rise as a power-law longward of 1.6$\mu$m \citep[e.g.][]{alonso06}.   These distinctly different SED shapes can, in principle, be exploited to diagnose the dominant power source of a galaxy \citep[e.g.][]{lacy04,stern05}, but do suffer from important ambiguities and limitations, in particular beyond $z\sim$ 2. In particular, red rest-frame NIR colors only indicate the presence of warm dust, and cannot discriminate between an AGN, with a  nuclear warm dust component and  an extreme infrared-luminous starburst whose warm dust is associated with HII regions \citep[e.g.][]{yun08}. Moreover, the separation of infrared-luminous galaxies into AGN and star-bursting systems is regardless an over-simplification, since the most luminous  ULIRGs at low and high redshifts  show evidence for both  AGN activity and intense star formation and  the optical/NIR emission will contain contributions from both processes. Nevertheless, an investigation of the rest-frame NIR-MIR SED shapes can provide basic clues to the nature of the galaxies residing within LABs.

\begin{figure}
\epsscale{1.1}
\plotone{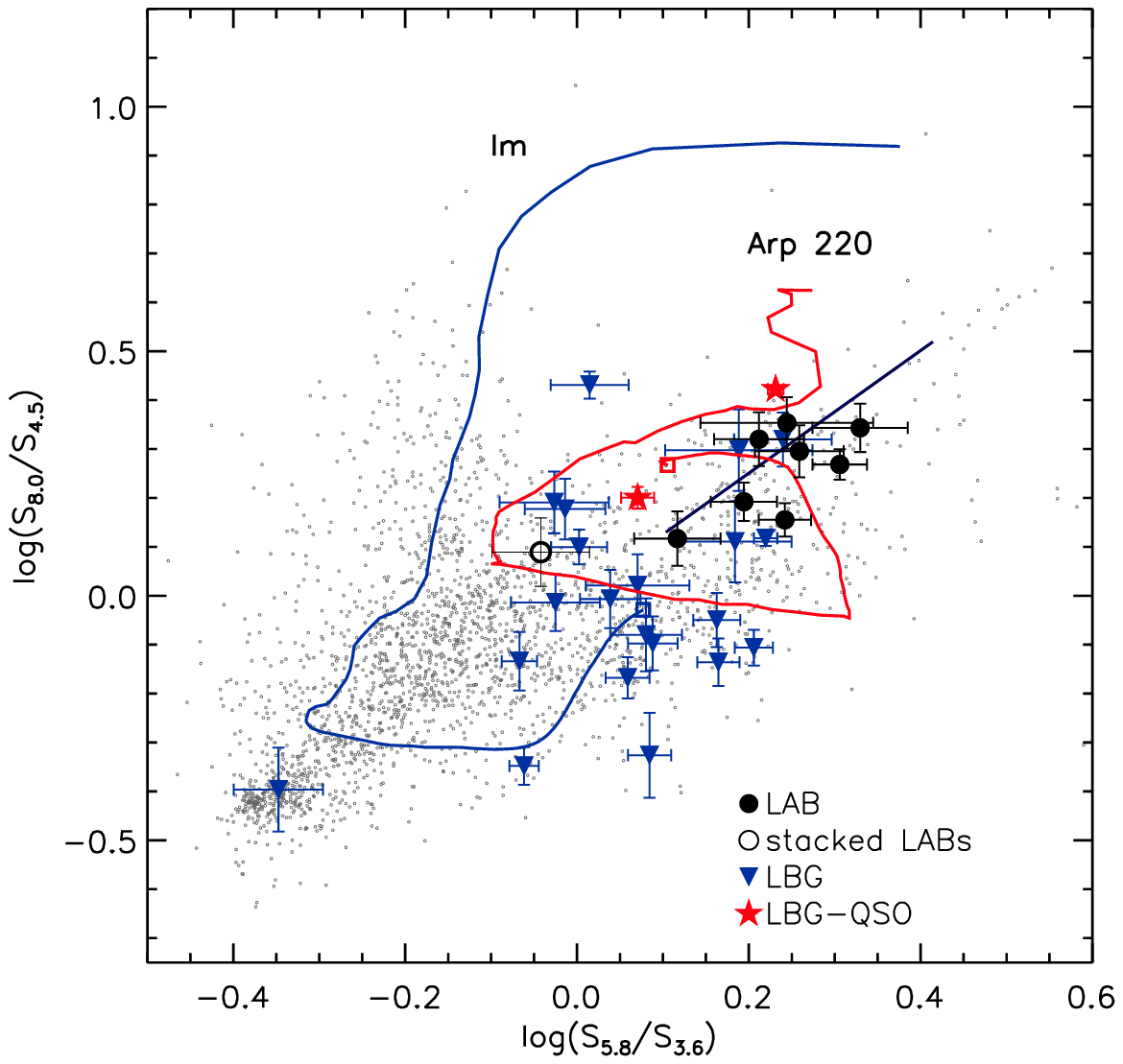}
\caption{The IRAC color-color diagram.   The grey points correspond to the entire 8$\mu$m-selected catalog in the extended SSA22 field ($>$10$\sigma$ in all 4 IRAC filters) and the filled black circles are those associated with the LAB counterparts ($>$5$\sigma$ in all 4 IRAC filters). Overlaid are two template SED tracks with redshift:  red denotes Arp220 and  blue shows an irregular  (Im) galaxy \citep{cww80}. Tracks run from $z$ = 0 to $z$ = 3.09 with the higher redshift end denoted by an open square.  
Also shown (black line) is the location of pure power-law galaxies ($S ~ {\propto} ~ \nu^\alpha$) with $\alpha$ = [-0.5,-2.0].  LBGs in the SSA22 field with 8$\mu$m detections are shown by the blue triangles; the two red stars are those LBGs confirmed as QSOs \citep{ste03}. The stacked detection of the LABs which are not individually detected at 8$\mu$m is shown by the open black circle.  
\label{cc}}
\end{figure}

In Figure \ref{cc} we show an IRAC color-color diagram for  LAB counterparts detected in all four IRAC filters \citep[e.g.][]{lacy04}. As discussed above, infrared power-law galaxies at low and high redshift inhabit a specific region of the diagram that is generally distinct from galaxies  dominated by stellar emission in the optical/infrared, but which overlaps with the location of  high-redshift star-forming ULIRGs. To illustrate this we plot the observed infrared colors of two representative SEDs with redshift, from $z=$ 0 to $z=$ 3.09: 
the ULIRG Arp220 and a typical Irregular starburst galaxy from \citet{cww80}.  Also shown are the observed infrared colors of 8$\mu$m-bright LBGs within the SSA22 structure \citep{ste03} and the general 8$\mu$m-selected population in the SSA22 field. 

  The LBGs exhibit  rest-frame optical/NIR colors spanning almost an order of magnitude; while the bulk of the objects have colors consistent with a $z\sim$ 3 blue starburst galaxy, a number extend into the region of the diagram inhabited by ULIRGs and AGN. The LAB counterparts on the other hand are found
without exception in the region of the redder LBGs  and AGN/ULIRG templates and are not well described by a low-extinction starburst.   In addition to the ULIRG/AGN ambiguity,  it is also difficult to constrain the redshifts using the observed NIR colors:  pure power-law systems in particular have a flat color-redshift relation.  Nevertheless, one can see from Figure \ref{cc} that among the 8$\mu$m-selected population  power-law (and power-law-like) galaxies  are rare ($\sim$10\%)  and thus the likelihood that all five LABs have a randomly aligned unassociated power-law galaxy (and two in the case of LAB01)  within the extent of the Lyman-$\alpha$ emission is remote ($\lesssim$ 1\% assuming a generous alignment distance of 5$\arcsec$), and we therefore conclude that all seven systems have $z = $ 3 and  are associated with their parent LAB.  

Although the majority of the LABs were not detected in all four IRAC filters, nor MIPS 24$\mu$m, we can still place constraints on their average properties through a stacking analysis of their infrared emission.  For this exercise we take as the source position the coordinates of the 3.6$\mu$m counterpart when available (the majority of the LABs) or if no counterpart exists at 3.6$\mu$m we simply take the peak of the Ly$\alpha$ emission;  only 1 LAB is too extended to identify a clear peak/center and has no clear counterpart  (LAB27), resulting in a sample of 20 systems, including LAB02-a.  
 We detect stacked emission at $>$ 5$\sigma$ in all four IRAC filters and  show the average color of the infrared-faint LAB systems in Figure \ref{cc}. As we see, they are not simply lower luminosity versions of the 8$\mu$m detected LABs, but rather are similar in color to the blue 8$\mu$m-detected LBG population, implying they are more moderate starburst systems. 

Returning to the infrared-bright LABs, the observed NIR colors 
require the presence of hot dust, heated by either  extreme  dust-enshrouded star formation or feedback from a central AGN.  As we now discuss however, the X-ray and MIR data indicate that AGN are present in some of these systems. First, turning to the X-ray results, two of the ten LBGs in this region are previously known optically confirmed QSOs \citep{ste03} which are strong X-ray emitters, and two of the seven LAB counterparts have weak X-ray emission (see \S 3).    Here we use the \citet{bau04}  AGN/starburst classification scheme where a  galaxy is considered an AGN if it satisfies any one of five X-ray or spectral conditions.   Two of these conditions are L$_X>$ 3$\times$10$^{42}$ erg/s and $f_{0.5-8.0 keV}/f_R > 0.1$  which are satisfied by the two X-ray detected LABs ( L$_X\sim$ 4$\times$10$^{43}$ erg/s and $f_{0.5-8.0 keV}/f_R \sim$ 1.5).    The individual and stacked X-ray limits for the remaining LABs are generous (L$_X<$ 8$\times$10$^{43}$ erg/s and L$_X<$ 6$\times$10$^{42}$ erg/s respectively) and are consistent with AGN according to these criteria.  Deeper X-ray studies (Geach et al. 2008, in preparation) will clarify this issue.

\begin{figure}
\epsscale{1.1}
\plotone{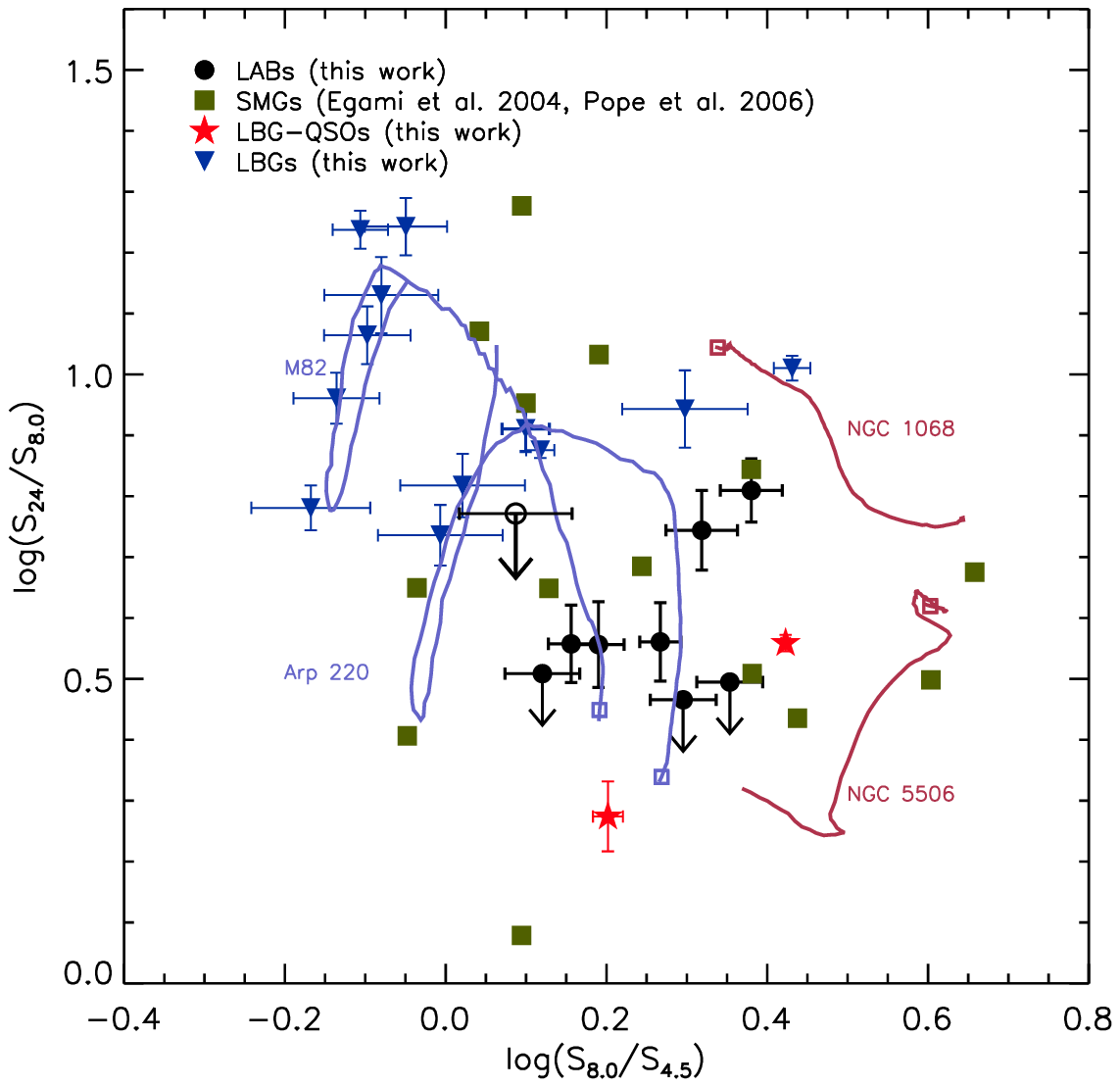}
\caption{An IRAC/MIPS color-color diagram, restricted to those objects detected in all four IRAC filters, as well as  at 24$\mu$m.   Filled black circles correspond to the LAB counterparts;  filled blue triangles denote the LBGs;  filled red stars show the LBGs which are confirmed QSO; and the solid green squares show SMGs with IRAC/MIPS detections which lie at $z\sim$ 3 from \citet{ega04} and \citet{pope06}. Also shown is the stacked detection of the infrared-faint LAB (open black circle).  Overlaid, to show the approximate distinction between the colors of AGN and ULIRGs are the template SEDs M82, Arp 220 (star-forming galaxies) and NGC 5506, NGC 1068 (dusty AGN).  \label{iracmips}}
\end{figure}

In Figure \ref{iracmips} we look at the constraints placed by the 24$\mu$m data through an investigation of the IRAC-MIPS colors.  Here we are limited to the LAB and LBG comparison systems which are also detected at 24$\mu$m. 
  AGN-dominated systems generally inhabit the lower right of this diagram: hot dust enhances their observed  8$\mu$m flux (2$\mu$m in the rest-frame) over the shorter wavelength stellar emission, but the lack of strong PAH features results in blue 24$\mu$m-8$\mu$m colors, with some variation due to the steepness of the power-law.  Star forming systems, on the other hand have less enhanced observed 8$\mu$m (rest-frame 2$\mu$) emission, and thus blue 8.0$\mu$m-4.5$\mu$m colors, but exhibit  a wide variety of PAH strengths and therefore a range in 24$\mu$m-8$\mu$m colors.   Such a diagram has been used by other authors to attempt to identify the power sources within the very luminous SMG population \citep[e.g][]{ega04,pope06}  which contain AGN and star-formation contributions, sometimes within the same objects, and, as seen in Figure \ref{iracmips}, tend to populate the entire color space. 
  
  We see a clear separation of AGN and star-forming systems within the 24$\mu$m-bright LBG population.  It is interesting to see that the LAB counterparts occupy a small region of the color-space, and transition  between these two populations.  Thus while smoking-gun evidence of AGN in these systems is lacking,      
  they appear distinct from the presumably pure star-forming LBG galaxies, but are also exhibit bluer NIR colors than AGN. We again perform a stacking analysis on the infrared faint LABs, this time restricting the sample to those which fall on the MIPS image. No MIPS flux is detected from these systems as a whole and we place an upper-limit on their observed 24$\mu$m/8$\mu$m color on Figure \ref{iracmips}. Here the separation between the infrared-bright and -faint LABs is less clear, though the infrared-faint LABs again lie blue-ward of the infrared-bright systems, toward the star-forming region of the diagram.

 \subsection{Agreement with Submillimeter Results}
\citet{gea05}  observed 25 of the 35 LABs from \citet{mat04} at 850$\mu$m and detected 5 at $>$ 3$\sigma$ significance. One of these, LAB10, lies outside of our imaged region. 
We find an almost one to one correspondence between the 8$\mu$m and 24$\mu$m detections presented here and the submillimeter detections.  We recover all 4  of the 850$\mu$m sources (for which we have data) at 8$\mu$m. At 24$\mu$m we fail to detect only LAB05 (limits are given in Table 1). 
 
The 24$\mu$m-850$\mu$m flux ratios or limits are in good agreement with those measured  for $z\sim$ 3 SMGs ($<$ 0.02) by \citet{pope06}.   As extensively discussed by \citet{pope06} such low values are consistent with galaxies such as Arp220 and Mrk273, which are not typical ULIRGs,  and indicate higher levels of extinction and/or cooler dust temperatures than generally seen for ULIRGs.   Cooler dust temperatures may be reached if the dust is spatially extended  compared to local compact ULIRGs.  This explanation is particularly interesting for LAB01: recent observations with the SMA \citep{mat07} have failed to detect the $S_{850{\mu}m} \sim$ 17$\mu$Jy source in the blob and this places a lower limit on its spatial extent of $>$ 4{\arcsec} in the submillimeter (assuming it is  signal dust enshrouded system).   Such an explanation however cannot account for the failure of recent ASTE-AzTec observations to detect this source with a 5$\sigma$ limit of $\sim$ 10mJy \citep{koh08}. 

As we showed in the previous section the 8$\mu$m flux of the 8$\mu$m-detected systems cannot be due only to stellar emission, but must be enhanced by the presence of hot dust,  associated with prodigious star formation or due to a central dust enshrouded AGN.  We have not detected any 8$\mu$m source within an LAB which is not also detected at 850$\mu$m, with the exception of LAB16, whose 850$\mu$m limit is 4$\times$ higher than the 4 other 850$\mu$m sources.  

Adopting the average 850$\mu$m-8$\mu$m color for the infrared-bright LABs, a non-detection at 850$\mu$m of $<$ 3mJy (the approximate depth of Geach et al.) would lie below our 8$\mu$m detection limit and therefore, based on this color alone, the infrared-faint LABs could simply be lower luminosity versions of the same kinds of systems.  A non-detection at 850$\mu$m \citep{gea05} and 24$\mu$m (this work) does not place very stringent limits on the bolometric luminosities of these systems however, and they remain compatible with ULIRG-level luminosities. 
Alternatively, the infrared-faint systems could be starburst galaxies with low extinction levels and correspondingly lower 850$\mu$m-8$\mu$m ratios. This scenario is in agreement with the stacked measurement on Figure \ref{cc} and with the additional color results discussed in the following sections.   We can rule out with certainty however,  dust enshrouded QSOs  of similar rest-frame 2$\mu$m luminosities, but which lack an extended cold dust component.

\subsection{The rest-frame UV-Optical-NIR properties}

\begin{figure}
 \epsscale{1.1}
\plotone{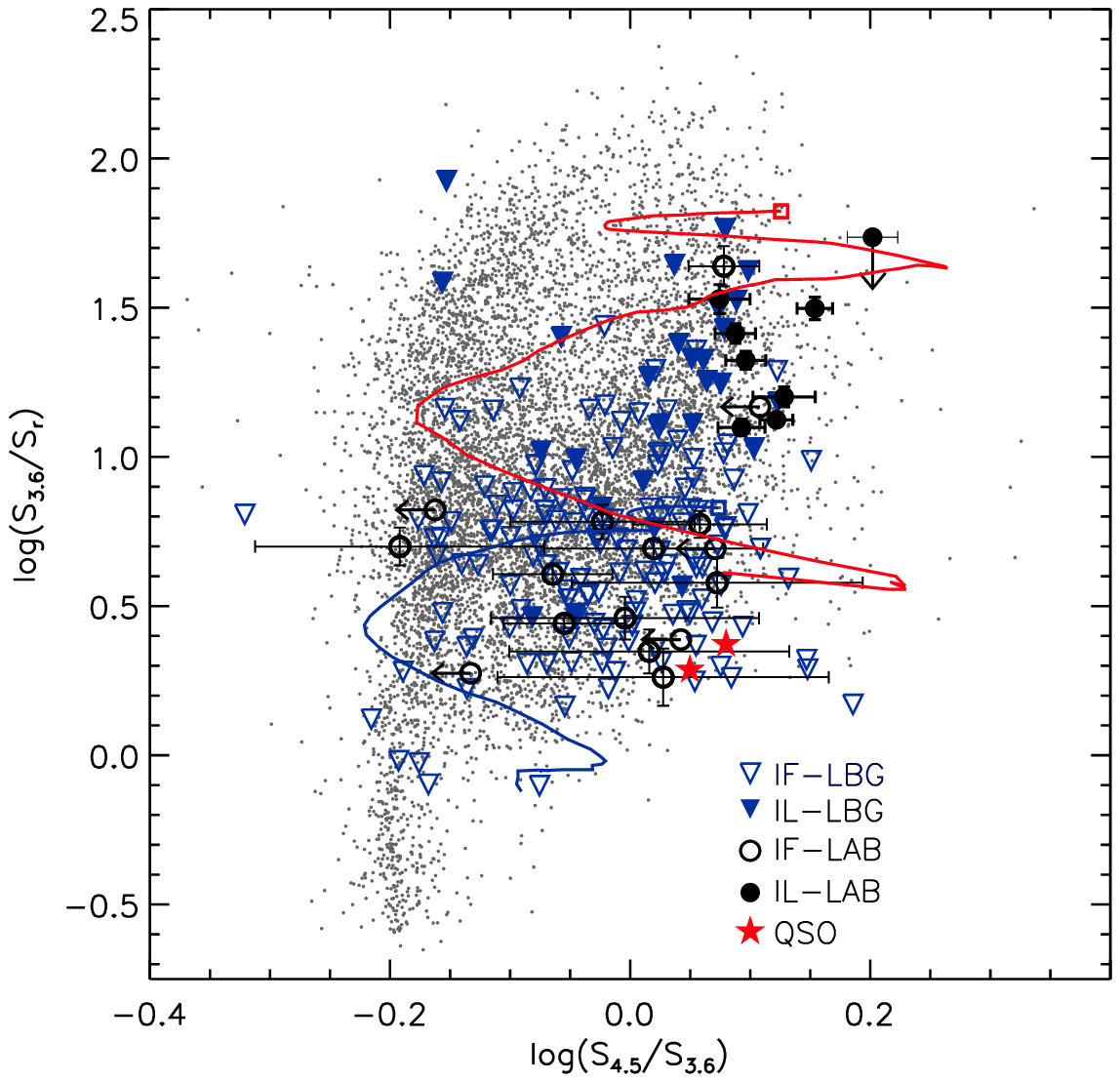}
\caption{ A rest-frame optical/UV color-color diagram. The small grey points correspond to the 3.6$\mu$m-selected population in the SSA22 field. Points are as defined in \ref{cc} and \ref{cm} with the exception of the open triangles (individual 8$\mu$m-faint LBGs) and the open circles (individual 8$\mu$m-faint LABs). Here we denote 8$\micron$-detected LABs and LBGs as infrared-luminous (IL) and 8$\micron$ non-detections as infrared-faint (IF).  Also shown are the SED tracks (from $z =$ 0 - 3.09)  for Arp 220 (red) and an irregular galaxy \citep{cww80} with the open square denoting $z =$ 3.09. \label{ccr}}
\end{figure}

The bulk of the LAB sample is not detected longward of 4.5$\mu$m, and in addition to the stacking analysis of the previous sections we can also look to the combination of 3.6$\mu$m and/or 4.5$\mu$m data with the NIR and optical imaging.   In Figure \ref{ccr} we show a rest-frame UV/optical/NIR color-color diagram for the LBGs and LABs.  Here again, we see a clear separation between 8$\mu$m-detected and non-detected systems, be they LABs or LBGs. The infrared-bright galaxies continue to show red colors into the rest-frame UV, indicating either older stellar populations and/or high levels of extinction.  Notably, as seen in Fig \ref{cc} the infrared-faint LABs are well-separated in color from the infrared-bright systems, and are very similar to the general infrared-faint LBGs.  The exceptions are the two infrared-bright LBG-QSOs, which show extremely blue UV/optical colors.
 
\begin{figure}
\epsscale{1.1}
\plotone{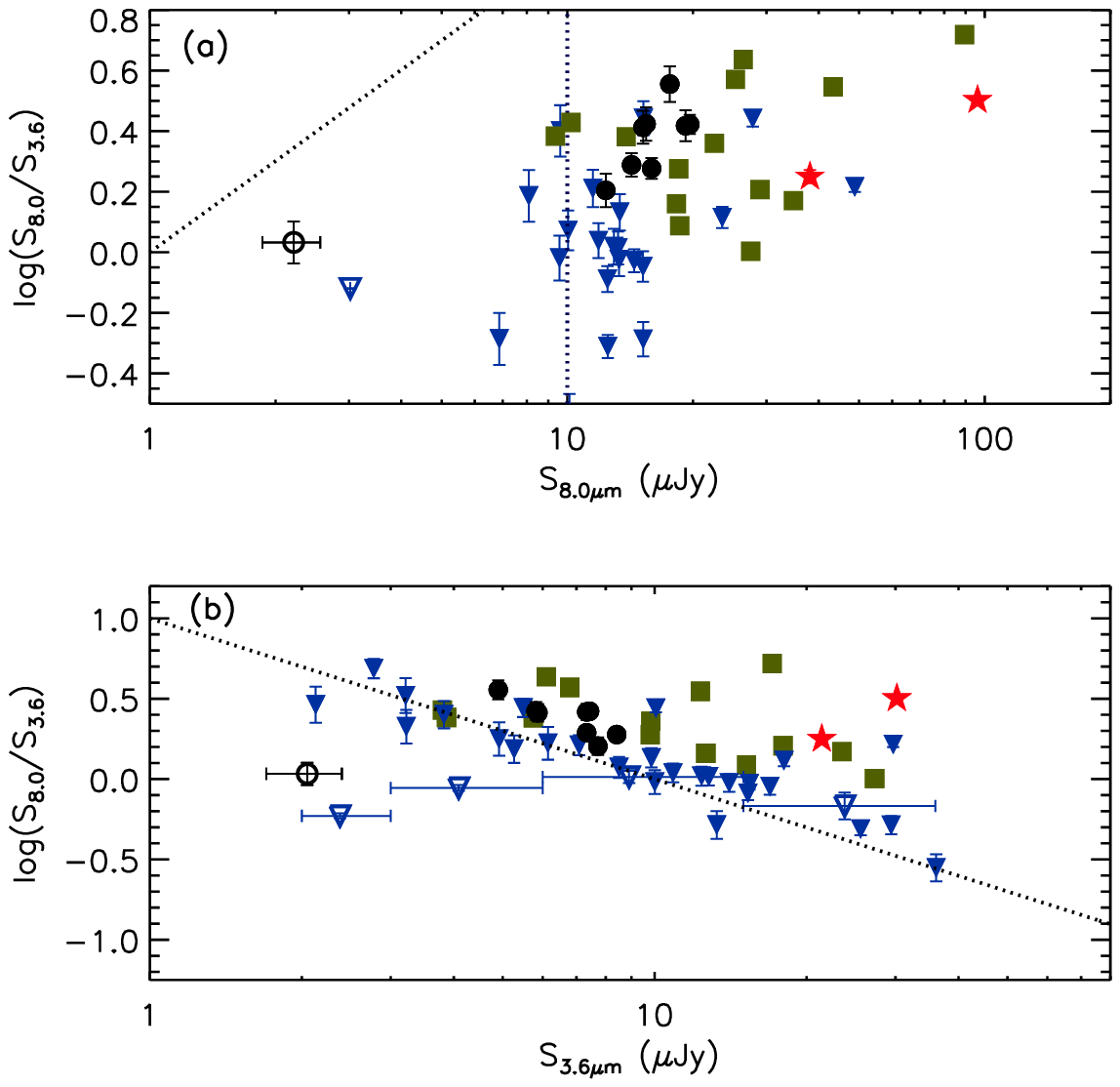}
\caption{(a) The optical/NIR color as a function of rest-frame $\sim$ 2$\mu$m.  The vertical dotted line denotes the approximate 5$\sigma$ limit of the 8$\mu$m data. The diagonal dotted line corresponds to the 8$\mu$m/3.6$\mu$m flux ratio, at the 3.6$\mu$m detection limit. Points are the same as in Figure 4. The filled black circles are the 8$\mu$m-detected LAB counterparts; the filled blue triangles are the 8$\mu$m-detected LBGs, with the two solid red stars corresponding to the confirmed QSOs, and the solid green squares are  SMGs \citep{ega04,pope06} for comparison. The open black circle and blue triangle  below the 8$\mu$m detection limit correspond to stacked measurements for the 8$\mu$m-faint LABs and LBGs respectively.  (b)    The optical/NIR color as a function of rest-frame $\sim$0.9$\mu$m. The dotted line shows the color limit at the 8$\mu$m depth. The points are as in (a), however the LBG stacked detections include both 8$\mu$m-faint  and  8$\mu$m-bright objects. \label{cm}}
\end{figure}

In Figure \ref{cm} we show the observed 8.0$\mu$m/3.6$\mu$m flux ratio as a function of observed 8.0$\mu$m and, alternately, 3.6$\mu$m flux.  Again we compare to the LBG population, and add to the analysis the SMGs and QSOs as in Figure \ref{iracmips}.  In Figure \ref{cm}(a) we see a correlation between the rest-frame 2.0$\mu$m and the rest-frame NIR/optical colors, in the sense that galaxies which are more luminous at 2.0$\mu$m exhibit redder colors, and the stacking analysis confirms that this relation continues to fainter levels.    
  Visually, the slope of the relation appears to lie between a flat color-luminosity relation and that of the line representing constant 3.6$\mu$m flux; that is, the relation results from an overall increase in luminosity and a enhancement of the 2.0$\mu$m flux over the rest-frame optical.    The rest-frame $\sim$ 2.0$\mu$m is often taken as a proxy for stellar mass, however we have already argued that this wavelength is contaminated by AGN or intense star-formation in many of these systems.  Indeed, the most luminous systems at observed 8.0$\mu$m are the confirmed QSOs, followed by the LABs and then the infrared-bright LBG population. Though clearly not a complete comparison sample this does again hint that the LABs are transition objects between pure AGN and star-formation dominated systems. 
  
  In Figure \ref{cm}(b) we plot the same color, but now against observed 3.6$\mu$m flux.   In addition to illustrating the spread in 3.6$\mu$m luminosities, this allows us to probe the observed 8$\mu$m flux of the comparison LBG population  as a function of 3.6$\mu$m flux.  The statistical detection of the 8$\mu$m-faint LABs fits within  the general behavior of the LBGs, which become markedly bluer in the optical/NIR with decreasing 3.6$\mu$m luminosity.  Thus,  the fainter, presumably less massive systems contain lower levels of dust, star formation and AGN activity.

\subsection{Ly$\alpha$ excitation mechanisms and implications for galaxy evolution within the SSA22 structure }

The two central questions concerning LABs are:  what drives the extended Ly$\alpha$ emission and, more generally, how are these structures related to galaxy formation?  There are two basic scenarios for the relationship between the LABs and their counterpart galaxies.  In the first, the embedded object is the excitation power source through processes such as photoionization by a strong UV emitter or by shock-heating due to super-galactic scale winds from a very  strong starburst or a central AGN.  In the second scenario, the Ly$\alpha$ emision could be due to heating through gravitational collapse; in such a picture, the LAB and counterpart have no causal relationship  (that is, the galaxy is not driving the Ly$\alpha$ emission), but rather both mark sites of galaxy formation activity. In either case we might expect a correlation between the properties of the counterpart and the Ly$\alpha$ halo; in the first case the correlation is obvious and causal while in the second case we might, for example, expect the most massive and luminous LABs to trace regions of correspondingly massive galaxy formation activity such as major-mergers, AGN, and extreme infrared starbursts.

Much work has been done to attempt to separate these different scenarios, but no common picture which works for all LABS has yet emerged.    M04 extensively discussed the possible excitation mechanisms of the 35 LABs in SSA22.  Using simple assumptions, such as a Salpeter IMF and an unhindered line of sight,  they showed that $\sim$ 1/3 of the LABs do not contain galaxies which are sufficiently UV-luminous to power the LABs. Of   the six 8$\mu$m-detected LABs discussed here,   four (LAB01, LAB02, LAB05, LAB14) are not sufficiently bright in the rest-frame UV  unless the bulk of the ionizing radiation is emitted away from our line of sight. 
This is not an unreasonable explanation since all six objects are dusty and IR-luminous, as evidenced by their 8$\mu$m, 24$\mu$m, and/or 850$\mu$m emission.  The remaining  LABs however, have UV/optical/IR colors which imply lower levels of dust extinction and are therefore unlikely to be highly obscured, but still lack obvious UV emitters.     

   By probing both hot and cold dust emission a number of authors \citep{gea05, col06,dey05} have shown that a small fraction of LABs host infrared-bright systems of ULIRG and HyLIRG  luminosity levels and importantly, that the inferred bolometric luminosity of these systems is weakly correlated with the Ly$\alpha$ luminosity, though roughly three orders of magnitude larger.  To the infrared-bright sample of \citet{gea05} we add LAB16;  although not detected at 850$\mu$m we may extrapolate from the 24$\mu$m flux (assuming a similar SED as the other infrared-bright LABs) and find that LAB16 follows the roughly same L$_{Ly\alpha}$-L$_{bol}$ relation. As argued by \citet{gea05}, this suggests that the process which heats the dust, i.e.~a powerful AGN or intense star formation, is also responsible for exciting the gas halo, and a natural mechanism for this is galaxy-scale winds; however, this relation could simply indicate a natural non-causal correlation between the size of the collapsing gas cloud and the galaxy formation activity  
 occurring within it. 
 
We can use the data presented here to add to these studies by directly probing the nature of the LAB counterparts, though as with the earlier work  this cannot prove a causal connection between the embedded galaxy and the gas emission. Throughout the paper however we have included similar analyses on the LBG population within the SSA22 structure, with the motivation that a comparative study of those systems which do not exhibit extended Ly$\alpha$ emission might further
elucidate the processes responsible for the LABs.  

We can divide the sample into two distinct groups:  those systems, be they LBGs or LABs, which are detected at 8$\mu$m, hereafter referred to as  infrared-luminous (IL), and galaxies which are not detected at 8$\mu$m,  or infrared-faint (IF).  Let us first discuss the IL sample. The MIR/NIR colors of the LBGs and LABs within this group are similar and are best described by either a dusty AGN power-law or an intense
starburst ULIRG.  Some of these have been detected in the X-ray with emission levels consistent with though not uniquely indicative of  AGN. Differences begin to arise when we look at the UV/optical colors or include the rest-frame 6$\mu$m in the analysis (Figures \ref{ccr} and \ref{iracmips}).   The bulk of the LBG population is systematically bluer in UV/optical color than the LABs, while in MIR color there is a clear continuum of colors between the bulk of the LBGs, the LABs and the LBG-QSOs.    Finally, in Figure \ref{cm} we see hints of a similar gradient in rest-frame 2$\mu$m luminosity, such that the LABs lie between the confirmed LBG-QSO and the remainder of the LBG population.  
This hints that, like the SMG population, IL-LABs may harbor deeply enshrouded AGN, in addition to intense star formation. 

The following toy-model might therefore be proposed for these systems. The most luminous LABs are powered through galaxy-scale superwinds driven by intense star formation and AGN activity. As the AGN grows it eventually becomes powerful enough to completely blow-out the gas cloud and enshrouding dust and emerges as a naked AGN, still luminous at the shorter IR wavelengths but no longer an LAB. It is not clear from these data how the IL-LBGs fit into this picture.  As suggested by \citet{dey05}, one might speculate that the brightest LBGs are older systems  whose AGN has turned off, hence their lower 8$\mu$m fluxes, but whether a LAB phase is common to all LBGs is not yet clear.  This scenario could however account for the variety of morphologies of the LABs and the location of the galaxies within them. Take for example the three largest LABs, LAB01, LAB02 and the LAB of \citet{dey05}:  all have likely AGN counterparts near the edge of the Ly$\alpha$ emission, and these may be systems caught in the process of emerging from their parent cloud.

One might posit that an AGN is required to power the intense Ly$\alpha$ emission,  and this is the root of the difference between the LBG without extended Ly$\alpha$ emission and LAB populations.  This seems at odds however, with the IF systems.   Though the data are sparse for these groups, we see no difference between the properties of the IF LAB counterparts and the IF LBGs: both appear to be pure star-forming systems with roughly similar masses (now using the 8$\mu$m luminosity as a proxy for mass). In these systems AGN are not driving the Ly$\alpha$ emission, nor can the presence of a Ly$\alpha$ halo for one group but not the other be explained by obvious differences in  their counterpart properties.   

It seems more likely therefore that the difference is in evolutionary phase over a cross-section of masses;  those galaxies with extended Ly$\alpha$ emission are in  earlier phases of formation than the older LBGs.  In this case the Ly$\alpha$ emission could be powered by superwinds, as discussed above, or may be unrelated to (but still correlated with) the embedded galaxies and driven instead by gravitational heating, as suggested, for example, by \citet{dey05} and \citet{smi07}.  Again, however, such a scenario would seem to imply that an LAB phase is common to all systems, unless secondary factors such as morphology play a role.

\subsection{Conclusions}
We have used mid-infrared imaging from Spitzer IRAC and MIPS 24$\mu$m to investigate the nature of a large sample of  LAB systems within the SSA22 filament at $z=$ 3.09.   We have detected 6/26 of the objects in all four IRAC filters, with four also detected with MIPS 24$\mu$m, and find excellent correspondence with the 850$\mu$m detections of \citet{gea05}.  By analyzing the rest-frame UV, optical, and near/mid/far-infrared colors, and archival X-ray imaging, we show that these six systems exhibit 
signs of AGN activity and/or intense and dusty star formation;  notably, they exhibit colors which bridge the star-forming LBG galaxies within the SSA22 structure and objects optically identified as QSOs. 
Through a stacking analysis see that the infrared faint LABs are markedly different than the infrared-luminous systems, and are consistent with blue star-forming galaxies, with no evidence of an AGN contribution. 
These results are in line with a model in which the LAB Ly$\alpha$ luminosity is powered by large-scale superwinds, driven by star-formation in the smaller  systems and a combination of AGN and extreme starbursts in the most massive. The presence of such a halo may be dictated by the evolutionary phase of the galaxy, such that the halo-less LBGs are older systems than the LABs. A scenario in which the LAB is the result of gravitational heating and the galactic activity is not driving the LAB luminosity cannot be ruled out, however.

\acknowledgments

Observations reported here were obtained at the MMT Observatory, a joint facility of the Smithsonian Institution and the University of Arizona.

{\it Facilities}: MMT, Spitzer, XMM, CXO


\begin{thebibliography}{}
\bibitem[Alonso-Herrero et al.(2006)]{alonso06} Alonso-Herrero, A. et al. 2006, ApJ, 640, 167
\bibitem[Basu-Zych \& Scharf (2004)]{basu04} Basu-Zych, A. \& Scharf, C. 2004, \apj, 615, L85
\bibitem[Bauer et al.(2004)]{bau04} Bauer, F.E., Alexander, D.M., Brandt, W.N., Schneider, D.P., Treister, e., Hornschemeier, A.E., \& Garmire, G.P. 2004, AJ, 128, 2048
\bibitem[Beleen et al.(2008)]{bel08} Beelen, A., et al. 2008, A\&A, 485, 645
\bibitem[Blum et al.(2006)]{blum06} Blum, R.D. et al. 2006, AJ, 132, 2034
\bibitem[Bower et al.(2004)]{bow04} Bower, R.G., et al. 2004, MNRAS, 351, 63
\bibitem[Champan et al.(2001)]{cha01} Chapman, S.~C., Lewis, G.~F., Scott, D., Richards, E., Borys, C., Steidel, C.~C., Adelberger, K.~L, \& Shapley, A.E. 2001, ApJ, 548, L17
\bibitem[Chapman et al.(2004)]{cha04} Chapman, S.~C., Smail, I., Windhorst, R., Muxlow, T., \& Ivison, R.J. 2004, 611, 732
\bibitem[Chapman et al.(2005)]{cha05} Chapman, S.C., Blain, A.W., Smail, I., Ivison, R.J. 2005, \apj, 622, 772
\bibitem[Cirasuolo et al.(2008)]{cir08} Cirasuolo, M., McLure, R.J., Dunlop, J.S., Almaini, O., Foucaud, S., \& Simpson, C. 2008, submitted to MNRAS, arXiv:0804.3471
\bibitem[Coil et al.(2004)]{coi04} Coil, A.L., Gerke, B.F., Newman, J.A., Ma, C.-P., Yan, R., Cooper, M.C., Davis, M., Faber, S., Guhathakarta, P., \& Koo, D.C. 2006, \apj, 638, 668
\bibitem[Colbert et al.(2006)]{col06} Colbert, J.~W., Teplitz, H., Francis, P., Palunas, P., Williger, G.~M., \& Woodgate, B. 2006, \apj, 637, L89
\bibitem[Coleman et al.(1980)]{cww80} Coleman, G.~D., Wu, C.~-C., \& Weedman, D.~W. 1980, ApJS, 43, 393
\bibitem[Dey et al.(2005)]{dey05} Dey, A. et al. 2005, \apj, 629, 654
\bibitem[Dey et al.(2008)]{dey08} Dey, A., et al. 2008, \apj, in press.
\bibitem[Egami et al.(2004)]{ega04} Egami, E., et al. 2004, ApJS, 154, 130
\bibitem[Fabbiano et al.(1989)]{fab89} Fabbiano, G. 1989, ARA\&A, 27, 87
\bibitem[Fazio et al.(2004)]{faz04} Fazio, G.G, et al. 2004, ApJS, 154, 10 
\bibitem[Fine et al.(2006)]{fin06} Fine, S., et al. 2006, MNRAS, 373, 613
\bibitem[Francis et al.(2001)]{fran01} Francis, P.J., et al. 2001, \apj, 554, 1001
\bibitem[Fynbo et al.(1999)]{fyn99} Fynbo, J.U., Moller, P., \& Warren, S.J. 1999, MNRAS, 305, 849
\bibitem[Geach et al.(2005)]{gea05} Geach, J.~E. et al. 2005, ApJ, 649, 661
\bibitem[Geach et al.(2007)]{gea07}  Geach, J.E., Smail, I., Chapman, S.C., Alexander, D.M., Blain, A.W., Stott, J.P., \& Ivison, R.J. 2007, \apj, 655, L9
\bibitem[Giavalisco et al.(1998)]{gia98} Giavalisco, M., Steidel, C.C., Adelberger, K.L., Dickinson, M.,E., Pettini, M., \& Kellogg, M. 1998, \apj, 503, 543 
\bibitem[Gordon et al.(2005)]{gor05} Gordon, K.~D. et al. 2005, PASP, 117, 503
\bibitem[Granato et al.(2006)]{gran06} Granato, G.~L., Silva, L., lapi, A., Shankar, F., De Zotti, G., \& Danese, L. 2006, \mnras, 368, L72
\bibitem[Hayashino et al.(2004)]{haya04} Hayashino, T. et al. 2004, AJ, 128, 2073
\bibitem[Kaiser (1984)]{kai84} Kaiser, N. 1984, \apj, 284, L9
\bibitem[Kohno et al.(2008)]{koh08} Kohno, K. et al. 2008, submitted
\bibitem[Keel et al.(1999)]{keel99} Keel, W.C., Cohen, S.H., Windhorst, R.A., Waddington, I. 1999, AJ, 118, 2547
\bibitem[Lacy et al.(2004)]{lacy04} Lacy, M. et al. 2004, ApJS, 154, 166
\bibitem[Lacy et al.(2007)]{lacy07} Lacy, M., Petric, A.O., Sajina, A., Canalizo, G., Storrie-Lombardi, L.J., Armus, L., Fadda, D., \& Marleau, F.R. 2007, AJ, 133, 186
\bibitem[Le F\`evre et al.(2005)]{lefev05} Le F\`evre, O. et al. 2005, A\&A., 439, L877
\bibitem[Matsuda et al.(2004)]{mat04} Matsuda, Y., et al. 2004, AJ, 128, 569 (M04)
\bibitem[Matsuda et al.(2005)]{mat05} Matsuda, Y. et al. 2005, \apj, 634, 125
\bibitem[Matsuda et al.(2007)]{mat07} Matsuda, Y., Iono, D., Ohta, K., Yamada, T., Kawabe, R., Hayashino, T., Peck, A.~B., \& Petitpas, G.~R. 2007, \apj, 667, 667
\bibitem[Palunas et al.(2004)]{pal04} Palunas, P., Teplitz, H.I., Francis, P.J., Williger, G.M., Woodgate, B.E. 2004, \apj, 602, 545
\bibitem[Patton et al.(2002)]{pat02} Patton, D.R. et al. 2002, \apj, 565, 208
\bibitem[Pope et al.(2006)]{pope06} Pope, A. et al. 2006, \mnras, 370, 1185
\bibitem[Prescott et al.(2008)]{pres08} Prescott, M.K.M., Kashikawa, N., Dey, A., Matsuda, Y. 2008, \apj, 678, L77
\bibitem[Reuland et al.(2003)]{reu03} Reuland, M. et al. 2003, \apj, 592, 755
\bibitem[Rieke et al.(2004)]{riek04} Rieke, G. et al. 2004, ApJS, 154, 25
\bibitem[Rudnick et al.(2006)]{rud06} Rudnick, G., et al. 2007, \apj, 650, 624
\bibitem[Schuster et al.(2006)]{schu06} Schuster, M.~T., Marengo, M., \& Patten, B.~M. 2006, SPIE, 6270, 65
\bibitem[Scott et al.(2004)]{scott04} Scott, J.~E., Kriss, G.~A., Brotherton, M., Green, R.~F., Hutchings, J., Shull, J.~M., \& Zheng, W. 2004, ApJ, 615, 135 
\bibitem[Shang et al.(2005)]{shang05} Shang, Z. et al. 2005, ApJ, 619, 41
\bibitem[Shapley et al.(2003)]{shap03} Shapley, A.E., Steidel, C.C., Pettini, M., Adelberger, K.L. 2003, \apj, 588, 65
\bibitem[Smith et al.(2007)]{smi07} Smith, D.J.B., \& Jarvis, M.J. 2007, MNRAS, 387, 49
\bibitem[Steidel et al.(1998)]{ste98} Steidel, C.~C., Adelberger, K.~L,  Dickinson, M., Giavalisco, M.,Pettini, M., \& Kellogg, M. 1998, \apj, 492, 428
\bibitem[Steidel et al.(2000)]{ste00} Steidel, C.C., Adelberger, K.L., Shapley, A.E., Pettini, M., Dickenson, M., Giavalisco, M. 2002, \apj, 523, 170
\bibitem[Steidel et al.(2003)]{ste03} Steidel C.~C.,  M., Adelberger, K.~L., Shapley, A.~E., Pettini, M., Dickinson, M., \& Giavalisco, M. 2003, \apj, 592, 728
\bibitem[Stern et al.(2005)]{stern05} Stern, D. et al. 2005, \apj, 631, 163
\bibitem[Webb et al.(2003)]{web03} Webb, T.M.A., Eales, S.A., Lilly, S.J., Clements, D.L., Dunne, L., Gear, W.K., Ivison, R.J., Flores, H., \& Yun, M. 2003, \apj, 587, 41  
\bibitem[Werner et al.(2004)]{wer04} Werner, M.~W. et al. ApJS, 154, 1
\bibitem[Yun et al.(2008)]{yun08} Yun, S. et al. 2008, MNRAS, in press

\end{thebibliography}
\end{document}